\documentclass[zpreprint,zbstdefault]{./LaTeX/zeus/zeus_paper}
\usepackage[english]{babel}
\usepackage{./LaTeX/styles/lineno}
\chardef\usc=95
\chardef\til=126
\catcode`\@=11 
\DeclareRobustCommand\xdotspace{\futurelet\@let@token\@xdotspace}
\def\@xdotspace{%
  \ifx\@let@token.\else
  \ifx\@let@token\bgroup.\else
  \ifx\@let@token\egroup.\else
  \ifx\@let@token\/.\else
  \ifx\@let@token\ .\else
  \ifx\@let@token~.\else
  \ifx\@let@token!.\else
  \ifx\@let@token,.\else
  \ifx\@let@token:.\else
  \ifx\@let@token;.\else
  \ifx\@let@token?.\else
  \ifx\@let@token/.\else
  \ifx\@let@token'.\else
  \ifx\@let@token).\else
  \ifx\@let@token-.\else
  \ifx\@let@token\@xobeysp.\else
  \ifx\@let@token\space.\else
  \ifx\@let@token\@sptoken.\else
   .\space
   \fi\fi\fi\fi\fi\fi\fi\fi\fi\fi\fi\fi\fi\fi\fi\fi\fi\fi}
\catcode`\@=12 

\newcommand{\stru}[2]{%
   \relax\ifmmode\hbox{\vrule height#1 depth#2 width0pt}%
   \else\vrule height#1 depth#2 width0pt\fi}

\newcommand{\Ronum}[1]{\uppercase\expandafter{\romannumeral#1}}
\newcommand{\ronum}[1]{\expandafter{\romannumeral#1}}
\DeclareRobustCommand{\LaTeXZ}{%
  \LaTeX\kern-.05em4\kern-.1em
  {\raisebox{-0.2ex}{$\scriptstyle\text{ZEUS}$}}\xspace}

\DeclareMathAlphabet{\mathbf}{OT1}{cmr}{bx}{sl}
\newcommand{\eVdist}{\kern-0.06667em}

\newcommand{\Gev}{{\text{Ge}\eVdist\text{V\/}}}

\newcommand{\gev}{{\,\text{Ge}\eVdist\text{V\/}}}

\newcommand{\Tesla}{\,\text{T}}

\newcommand{\slashfrac}[2]{%
  \raisebox{0.5ex}{\ensuremath #1}\kern-0.12em/\kern-0.08em
  \raisebox{-.8ex}{\ensuremath #2}}

\newcommand{\sqr}[3]{%
    {\vcenter{\hrule height.#3ex\hbox{\vrule width.#2ex height#1ex
     \kern#1ex\vrule width.#3ex}\hrule height.#2ex}}}

\newcommand{\widebar}[1]{%
   \mkern1.5mu\overline{\mkern-1.5mu#1\mkern-1.mu}\mkern1.mu}
\catcode`\@=11 
\newcommand{\parenbar}{\mathpalette\p@renb@r}
\def\p@renb@r#1#2{\vbox{%
  \ifx#1\scriptscriptstyle \dimen@.7em\dimen@ii.2em\else
  \ifx#1\scriptstyle \dimen@.8em\dimen@ii.25em\else
  \dimen@1em\dimen@ii.4em\fi\fi \offinterlineskip
  \ialign{\hfill##\hfill\cr
    \vbox{\hrule width\dimen@ii}\cr
    \noalign{\vskip-.3ex}%
    \hbox to\dimen@{$\mathchar300\hfil\mathchar301$}\cr
    \noalign{\vskip-.3ex}%
    $#1#2$\cr}}}
\catcode`\@=12 

\newcommand{\pbar}{\widebar{p}}

\newcommand{\qbar}{\widebar{q}}

\newcommand{\cbar}{\widebar{c}}
\newcommand{\bbar}{\widebar{b}}

\newcommand{\IP}{{\rm I$\kern-0.01667em$P}\xspace}

\mathchardef\qsm=63
\mathchardef\pls=43
\mathchardef\mns=512
\mathchardef\plm=518
\mathchardef\eql=61
\mathchardef\smallleft=300
\mathchardef\smallright=301
\mathchardef\les=316
\mathchardef\gre=318
\mathchardef\leq=532
\mathchardef\grq=533

\catcode`\@=11 
\newcounter{pict@width}
\newcounter{pict@height}
\newlength{\pict@scale}
\newcommand{\psfigadd}[4]{%
\setcounter{pict@width}{1*\ratio{#2+\pict@scale/2}{\pict@scale}}
\setcounter{pict@height}{1*\ratio{#3+\pict@scale/2}{\pict@scale}}
\setlength{\unitlength}{\pict@scale}
\hbox to #2{\hspace{-\fill}\begin{picture}(\thepict@width,\thepict@height)
\put(0,0){\psfig{figure=#1,width=#2,height=#3,clip=}}
\SetScale{0.283466457}
\SetWidth{1.763889}
{#4}
\end{picture}}
}
\newcounter{pict@widthfst}
\newcounter{pict@widthscd}
\newcounter{pict@widthtot}
\newcommand{\psfigaddtwo}[7]{%
\setcounter{pict@widthfst}{1*\ratio{#2+\pict@scale/2}{\pict@scale}}
\setcounter{pict@widthscd}{1*\ratio{#2+#4+\pict@scale/2}{\pict@scale}}
\setcounter{pict@widthtot}{1*\ratio{#2+#4+#6+\pict@scale/2}{\pict@scale}}
\setcounter{pict@height}{1*\ratio{#3+\pict@scale/2}{\pict@scale}}
\setlength{\unitlength}{\pict@scale}
\hbox{\hspace{-\fill}\begin{picture}(\thepict@widthtot,\thepict@height)
\put(0,0){\psfig{figure=#1,width=#2,height=#3,clip=}}
\put(\thepict@widthscd,0){\psfig{figure=#5,width=#6,height=#3,clip=}}
\SetScale{0.283466457}
\SetWidth{1.763889}
{#7}
\end{picture}}
}
\newcommand{\psfigror}[4]{%
\setcounter{pict@width}{1*\ratio{#2+\pict@scale/2}{\pict@scale}}
\setcounter{pict@height}{1*\ratio{#3+\pict@scale/2}{\pict@scale}}
\setlength{\unitlength}{\pict@scale}
\hbox{\begin{picture}(\thepict@width,\thepict@height)
\put(0,\thepict@height){\psfig{figure=#1,width=#3,height=#2,clip=,angle=270}}
\SetScale{0.283466457}
\SetWidth{1.763889}
{#4}
\end{picture}}
}
\newcommand{\psfigrol}[4]{%
\setcounter{pict@width}{1*\ratio{#2+\pict@scale/2}{\pict@scale}}
\setcounter{pict@height}{1*\ratio{#3+\pict@scale/2}{\pict@scale}}
\setlength{\unitlength}{\pict@scale}
\hbox{\begin{picture}(\thepict@width,\thepict@height)
\put(0,0){\psfig{figure=#1,width=#3,height=#2,clip=,angle=90}}
\SetScale{0.283466457}
\SetWidth{1.763889}
{#4}
\end{picture}}
}
\catcode`\@=12 
\newlength\listtextwidth

\catcode`\@=11 
\newlength{\@tabfninsert}
\newlength{\@tabfnwidth}
\newcommand{\tabfootnote}[2]{%
  \setlength{\@tabfninsert}{0.8em}
  \setlength{\@tabfnwidth}{\textwidth}
  \addtolength{\@tabfnwidth}{-\@tabfninsert}
  \addtolength{\@tabfnwidth}{-0.4em}
  \noindent\makebox[\@tabfninsert][r]{\footnotesize$^{#1}$\hfil}\hfill%
  \parbox[t]{\@tabfnwidth}{\footnotesize #2\hfill}}
\catcode`\@=12 

\def\citeCTD{{\cite{%
nim:a279:290,*npps:b32:181,*nim:a338:254%
}}\xspace}
\def\citeMVD{{\cite{%
nim:a581:656%
}}\xspace}
\def\citeCAL{{\cite{%
nim:a309:77,*nim:a309:101,*nim:a321:356,*nim:a336:23%
}}\xspace}

\begin{document}
\title{
\vspace{-5cm}
\begin{flushright} {\normalsize \tt DESY 08-210}\\ \vspace{-.25cm}{\normalsize 
\tt December 2008} \end{flushright}
\vspace{2cm}
Measurement of beauty photoproduction using decays into muons in dijet events at HERA
}                                                       
                    
\author{ZEUS Collaboration}
\draftversion{9.0}
\date{21.\ December 2008}

\abstract{
Beauty photoproduction in dijet events has been measured at HERA with the ZEUS detector using an integrated luminosity of 126 pb$^{-1}$. Beauty was identified in events with a muon in the final state by using the transverse momentum of the muon relative to the closest jet. Lifetime information from the silicon vertex detector was also used; the impact parameter of the muon with respect to the primary vertex was exploited to discriminate between signal and background. Cross sections for beauty production as a function of the muon and the jet variables as well as dijet correlations are compared to QCD predictions and to previous measurements. The data are well described by predictions from next-to-leading-order QCD.

}

\makezeustitle

\def\3{\ss}                                                                                        
\pagenumbering{Roman}                                                                              
                                                   %
\begin{center}                                                                                     
{                      \Large  The ZEUS Collaboration              }                               
\end{center}                                                                                       
  S.~Chekanov,                                                                                     
  M.~Derrick,                                                                                      
  S.~Magill,                                                                                       
  B.~Musgrave,                                                                                     
  D.~Nicholass$^{   1}$,                                                                           
  \mbox{J.~Repond},                                                                                
  R.~Yoshida\\                                                                                     
 {\it Argonne National Laboratory, Argonne, Illinois 60439-4815, USA}~$^{n}$                       
\par \filbreak                                                                                     
  M.C.K.~Mattingly \\                                                                              
 {\it Andrews University, Berrien Springs, Michigan 49104-0380, USA}                               
\par \filbreak                                                                                     
  P.~Antonioli,                                                                                    
  G.~Bari,                                                                                         
  L.~Bellagamba,                                                                                   
  D.~Boscherini,                                                                                   
  A.~Bruni,                                                                                        
  G.~Bruni,                                                                                        
  F.~Cindolo,                                                                                      
  M.~Corradi,                                                                                      
\mbox{G.~Iacobucci},                                                                               
  A.~Margotti,                                                                                     
  R.~Nania,                                                                                        
  A.~Polini\\                                                                                      
  {\it INFN Bologna, Bologna, Italy}~$^{e}$
% ...
\newpage                                                                                           
\enlargethispage{5cm}                                                                              
$^{\    1}$ also affiliated with University College London,                                        
United Kingdom\\                                                                                   
% ...
$^{\    5}$ supported by the research grant no. 1 P03B 04529 (2005-2008) \\                        
% ...
%
                                                           %
                                                           %
\begin{tabular}[h]{rp{14cm}}                                                                       
$^{a}$ &  supported by the Natural Sciences and Engineering Research Council of Canada (NSERC) \\  
$^{b}$ &  supported by the German Federal Ministry for Education and Research (BMBF), under        
          contract numbers 05 HZ6PDA, 05 HZ6GUA, 05 HZ6VFA and 05 HZ4KHA\\                         
$^{c}$ &  supported in part by the MINERVA Gesellschaft f\"ur Forschung GmbH, the Israel Science   
          Foundation (grant no. 293/02-11.2) and the U.S.-Israel Binational Science Foundation \\  
$^{d}$ &  supported by the Israel Science Foundation\\                                             
$^{e}$ &  supported by the Italian National Institute for Nuclear Physics (INFN) \\                
$^{f}$ &  supported by the Japanese Ministry of Education, Culture, Sports, Science and Technology 
          (MEXT) and its grants for Scientific Research\\                                          
$^{g}$ &  supported by the Korean Ministry of Education and Korea Science and Engineering          
          Foundation\\                                                                             
$^{h}$ &  supported by the Netherlands Foundation for Research on Matter (FOM)\\                   
$^{i}$ &  supported by the Polish State Committee for Scientific Research, project no.             
          DESY/256/2006 - 154/DES/2006/03\\                                                        
$^{j}$ &  partially supported by the German Federal Ministry for Education and Research (BMBF)\\   
$^{k}$ &  supported by RF Presidential grant N 1456.2008.2 for the leading                         
          scientific schools and by the Russian Ministry of Education and Science through its      
          grant for Scientific Research on High Energy Physics\\                                   
$^{l}$ &  supported by the Spanish Ministry of Education and Science through funds provided by     
          CICYT\\                                                                                  
$^{m}$ &  supported by the Science and Technology Facilities Council, UK\\                         
$^{n}$ &  supported by the US Department of Energy\\                                               
$^{o}$ &  supported by the US National Science Foundation. Any opinion,                            
findings and conclusions or recommendations expressed in this material                             
are those of the authors and do not necessarily reflect the views of the                           
National Science Foundation.\\                                                                     
$^{p}$ &  supported by the Polish Ministry of Science and Higher Education                         
as a scientific project (2006-2008)\\                                                              
$^{q}$ &  supported by FNRS and its associated funds (IISN and FRIA) and by an Inter-University    
          Attraction Poles Programme subsidised by the Belgian Federal Science Policy Office\\     
$^{r}$ &  supported by an FRGS grant from the Malaysian government\\                               
\end{tabular}                                                                                      
                                                           %
                                                           %

\pagenumbering{arabic} 
\section{Introduction}
\label{sec-int}
The production of beauty quarks in $ep$ collisions should be accurately calculable in perturbative Quantum Chromodynamics (QCD) since the large mass of the $b$ quark provides a hard scale. Therefore it is interesting to compare such predictions to results using photoproduction events where a low-virtuality photon, emitted by the incoming lepton, collides with a parton from the incoming proton. Previous photoproduction analyses presented by ZEUS used semi-leptonic decays into muons \cite{pr:d70:012008,desy-08-129} and electrons \cite{epj:c18:625,pr:d78:072001}, and found agreement with next-to-leading-order (NLO) QCD calculations and predictions based on the $k_T$ factorisation approach \cite{pr:d73:114018}. Leptons from beauty decays were distinguished from other decay leptons and background by their large transverse momentum, $p_{T}^{\rm \: rel}$, relative to the axis of the jet with which they are associated.

For the analysis presented here, events with at least two jets ($jj$) were selected and $b$ quarks were identified through their decay into muons with large $p_{T}^{\rm \: rel}$ and large impact parameter, $\delta$, defined as the distance of closest approach of the muon with respect to the beam position. The latter was facilitated by the ZEUS silicon microvertex detector (MVD)\citeMVD. The impact parameter is large for muons from $b$ decays due to the long lifetime of $B$ hadrons. A combination of the $p_{T}^{\rm \: rel}$ and $\delta$ methods was also used by the H1 collaboration \cite{epj:c41:453} and good agreement was found with both the ZEUS results and the NLO QCD prediction, except for an excess at low $p_{T}$ of the muon ($p_{T}^\mu$) and jet ($p_{T}^{j}$). The measurement presented here covers a kinematic region extending to lower $p_{T}^\mu$ than the previous ZEUS and H1 jet measurements \cite{pr:d70:012008,epj:c41:453}. The cross section for beauty production has also been measured in $p\pbar$ collisions at the S$p\pbar$S \cite{beautyUA10,*beautyUA1a,*beautyUA1,*beautyUA1b} and Tevatron colliders \cite{beautyCDF1,*beautyCDF2,*beautyCDF3,*beautyCDF4,*beautyCDF4a,*beautyCDF5,*beautyCDF5a,*beautyCDF7,*beautyCDF8,*beautyCDF9,*beautyD00,*beautyD01,*beautyD02,*beautyD03} and in $\gamma \gamma$ interactions at LEP \cite{beautyLEP1,*beautyLEP3,beautyLEP4}. Most results are in good agreement with QCD predictions but large discrepancies are observed in some \cite{beautyLEP1} of the results from $\gamma\gamma$ interactions at LEP.
				   
The dijet sample of beauty photoproduction events was also used to study higher-order QCD topologies. At leading order (LO), the two jets in the event are produced back-to-back in azimuthal angle, such that $\Delta\phi^{jj} =  \phi^{j1} - \phi^{j2} = \pi$. Additional soft radiation causes small azimuthal decorrelations, whilst $\Delta\phi^{jj}$ significantly lower than $\pi$ is evidence of additional hard radiation. Dijet correlations have been previously measured at ZEUS in inclusive-jet and charm photoproduction at high transverse energies \cite{np:b729:492,pr:d76:072011,pl:b565:87}; the conclusions from both are the same. Deviations from the NLO QCD predictions were found, especially in regions which are expected to be particularly sensitive to higher-order effects, i.e. at low $\Delta\phi^{jj}$. In this paper, the cross section versus $\Delta\phi^{jj}$ is presented for beauty photoproduction. These and other cross sections are compared to NLO QCD predictions and Monte Carlo models.

\section{Experimental set-up}
\label{sec-exp}

The analysis was performed with data taken in 2005 when HERA collided electrons with energy $E_{e} = 27.5 \, \rm GeV$ with protons of energy $E_{p} = 920 \, \rm GeV$, resulting in a centre-of-mass energy of $\sqrt{s} = 318 \, \rm GeV$. The results are based on an $e^{-}p$ sample corresponding to an integrated luminosity of 125.9 $\pm$ 3.3  pb$^{-1}$.

A detailed description of the ZEUS detector can be found elsewhere \cite{zeus:1993:bluebook}. A brief outline of the components that are most relevant for this analysis is given below.

In the kinematic range of the analysis, charged particles were tracked in the central tracking detector (CTD)~\citeCTD and the MVD~\citeMVD. These components operated in a magnetic field of $1.43\Tesla$ provided by a thin superconducting solenoid. The CTD consisted of 72~cylindrical drift chamber layers, organised in nine superlayers covering the polar-angle\footnote{The ZEUS coordinate system is a right-handed Cartesian system, with the $Z$ axis pointing in the proton beam direction, referred to as the``forward direction'', and the $X$ axis pointing left towards the centre of HERA. The coordinate origin is at the nominal interaction point.} region \mbox{$15^\circ<\theta<164^\circ$}. The MVD consisted of a barrel (BMVD) and a forward (FMVD) section with three cylindrical layers and four vertical planes of single-sided silicon strip sensors in the BMVD and FMVD respectively. The BMVD provided polar-angle coverage for tracks with three measurements from $30^\circ$ to $150^\circ$. The FMVD extended the polar-angle coverage in the forward region to $7^\circ$. After alignment the single-hit resolution of the BMVD was $25\,\mu \rm m$ and the impact-parameter resolution of the CTD-BMVD system for high-momentum tracks was $100\,\mu \rm m$.

The high-resolution uranium--scintillator calorimeter (CAL)~\citeCAL consisted of three parts: the forward (FCAL), the barrel (BCAL) and the rear (RCAL) calorimeters. Each part was subdivided transversely into towers and longitudinally into one electromagnetic section (EMC) and either one (in RCAL) or two (in BCAL and FCAL) hadronic sections (HAC). The smallest subdivision of the calorimeter was called a cell.  The CAL energy resolutions, as measured under test-beam conditions, were $\sigma(E)/E=0.18/\sqrt{E}$ for electrons and $\sigma(E)/E=0.35/\sqrt{E}$ for hadrons ($E$ in $\Gev$).

The muon system consisted of rear, barrel (R/BMUON) and forward (FMUON) tracking detectors. The B/RMUON consisted of limited-streamer (LS) tube chambers placed behind the BCAL (RCAL), inside and outside a magnetised iron yoke surrounding the CAL. The barrel and rear muon chambers cover polar angles from $34^\circ$ to $135^\circ$ and from $135^\circ$ to $171^\circ$, respectively. The muon system exploited the magnetic field of the iron yoke and, in the forward direction, of two iron toroids magnetised to $\sim 1.6$ T to provide an independent measurement of the muon momentum.

The luminosity was measured using the Bethe-Heitler reaction $ep \rightarrow e \gamma p$ by a luminosity detector which consisted of independent lead--scintillator calorimeter \cite{desy-92-066,*zfp:c63:391,*acpp:b32:2025} and magnetic spectrometer \cite{nim:a565:572} systems. The fractional systematic uncertainty on the measured luminosity was 2.6\%.

\section{Data selection}
\label{sec:dat}

The data were preselected by the ZEUS online trigger system \cite{zeus:1993:bluebook,nim:a580:1257,uproc:chep:1992:222} to contain two high-$p_{T}$ jets and/or a muon candidate.

The hadronic system (including the muon) was reconstructed offline from energy-flow objects (EFOs) \cite{thesis:briskin:1998} which combine the information from calorimetry and tracking and which were corrected for dead material and for the presence of muons. Jets were reconstructed offline from EFOs using the $k_T$ algorithm \cite{np:b406:187} in the longitudinally invariant mode \cite{pr:d48:3160}. The $E$-recombination scheme, which produces massive jets whose four-momenta are the sum of the four-momenta of the clustered objects, was used.

Muons were reconstructed by matching a track found in the CTD and the MVD with a track found in the inner chambers of the B/RMUON. Muons were associated with jets using the $k_{T}$ algorithm; if the EFO corresponding to a reconstructed muon was included in a jet then the muon was considered to be associated with the jet, which will from now on be referred to as the muon-jet.

Events with one muon and two jets were selected by requiring:
\begin{itemize}
\item $\ge 1$ muon with pseudorapidity $-1.6  <  \eta^\mu < 1.3$, and transverse momentum $p_{T}^\mu>2.5 \gev$ (this cut was lowered to $p_{T}^\mu>1.5 \gev$ for the measurement of the differential cross section with respect to $p_{T}^\mu$); the muon track was required to have at least 4 MVD hits;
\item $\ge 2$ jets with pseudorapidity $|\eta^{j}| < 2.5$, and transverse momentum $p_{T}^{j}>7 \gev$ for the highest-$p^{j}_{T}$ jet and $p_{T}^{j}>6 \gev$ for the second-highest-$p^{j}_{T}$ jet;
\item that the muon was associated with a jet with $p_{T}^{j}>6 \gev$ which is not necessarily one of the two highest-$p_{T}$ jets. To ensure a reliable $p_{T}^{\rm \: rel}$ measurement (see Section \ref{sec:cros}), the residual jet transverse momentum, calculated excluding the associated muon, was required to be greater than $2  \gev$;
\item no scattered-electron candidate \cite{nim:a365:508} found in the CAL;
\item $0.2<y_{\rm JB}<0.8$, where $y_{\rm JB}$ is the estimator of the inelasticity, $y$, measured from the EFOs according to the Jacquet-Blondel method \cite{proc:epfacility:1979:391}.
\end{itemize}
The last two cuts suppress the contributions from neutral current deep inelastic scattering events and from non-$ep$ interactions. The total efficiency of all these selection cuts was $\sim$23$\%$. A sample of 7351 events remained for $p_{T}^\mu>2.5 \gev$ and 14172 events remained for $p_{T}^\mu>1.5 \gev$.

\section{Signal extraction}
\label{sec:cros}

To evaluate detector acceptance and to provide the signal and background distributions, Monte Carlo (MC) samples of beauty, charm and light-flavour (LF) events were generated using {\sc Pythia} 6.2 \cite{cpc:135:238,epj:c17:137,hep-ph-0108264}, corresponding respectively to 9, 4.5 and 1 times the luminosity of the data. 
The generated events were passed through a full simulation of the ZEUS detector based on {\sc Geant} 3.21~\cite{tech:cern-dd-ee-84-1}. They were then subjected to the same trigger requirements and processed by the same reconstruction programs as the data. 

\begin{figure}[p]
\vfill
\begin{center}
\includegraphics[width=\textwidth]{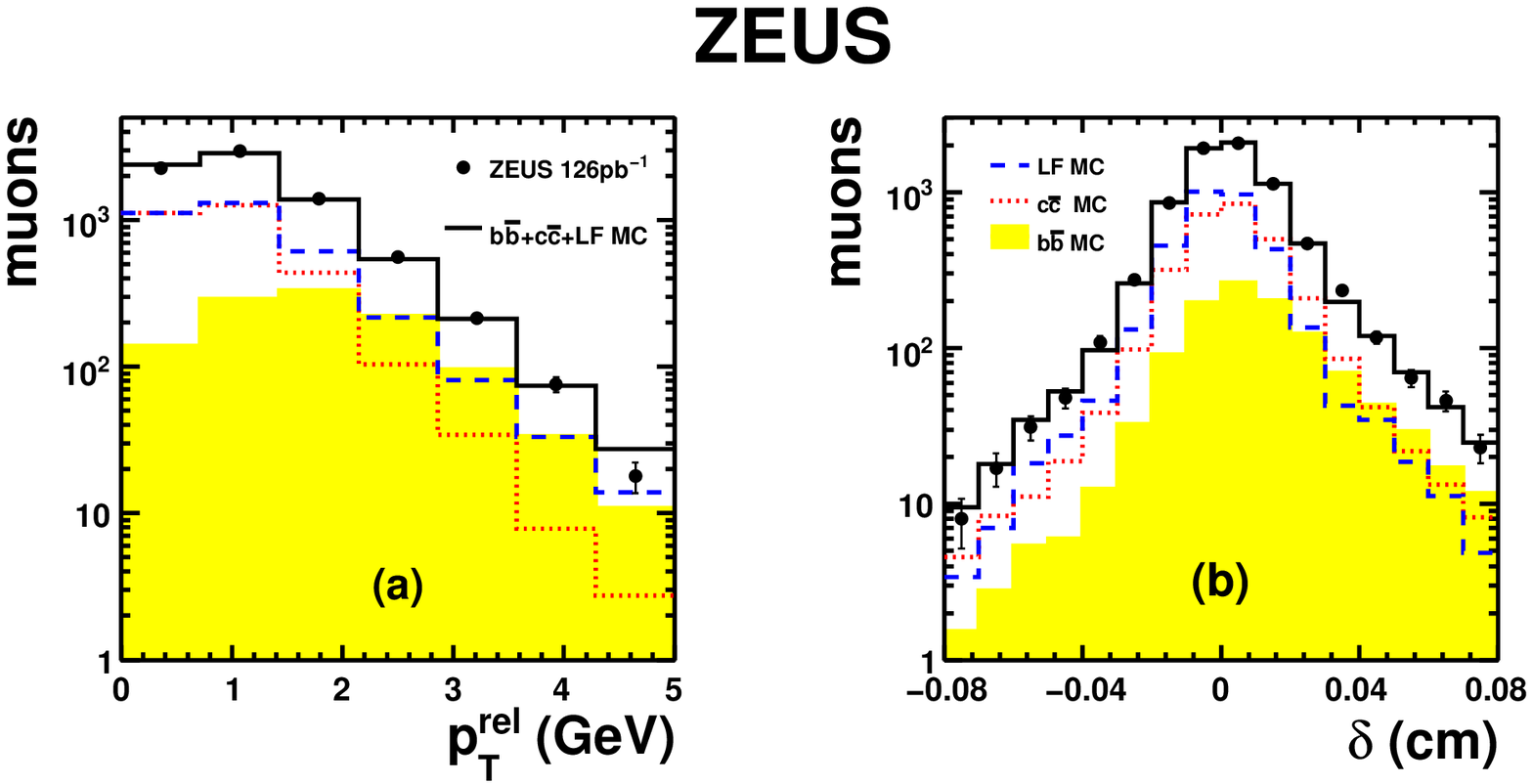}
\end{center}
\caption{
Distributions of (a) $p_{T}^{\rm \:rel}$ and (b) $\delta$. The data are compared to a mixture of beauty (shaded histogram), charm (dotted line) and light flavour(dashed line) {\sc Pythia} MC samples, combined with the fractions given by the two-dimensional $p_{T}^{\rm \:rel}$-$\delta$ fit. The total MC distribution is shown as the solid line.}
\label{f:ptrelip}
\vfill
\end{figure}

Due to the large $b$-quark mass, muons from semi-leptonic beauty decays tend to be produced with high transverse momentum with respect to the direction of the jet containing the $B$ hadron. The $p_{T}^{\rm \: rel}$ variable can therefore be exploited to extract the beauty signal; it is defined as:

\begin{equation}
p_{T}^{\rm \: rel}=\frac{ | \mathbf{p}^{\:\mu} \times ( \mathbf{p}^{j}-\mathbf{p}^{\:\mu} ) |}
          { | \mathbf{p}^{j}-\mathbf{p}^{\:\mu} | },
\end{equation}
where $\mathbf{p}^{\:\mu}$ is the muon and $\mathbf{p}^{j}$ the jet momentum vector. An underestimation of the tails of the $p_{T}^{\rm \: rel}$ distribution in the background MC was corrected as described in a previous publication \cite{pr:d70:012008}. The correction is applied in bins of $p_{T}^{\rm \: rel}$ to the LF MC sample. Half of the correction was also applied to the charm MC sample. The $p_{T}^{\rm \: rel}$ distribution for the selected data sample is compared to the MC simulation in Fig. \ref{f:ptrelip}(a). 

Muons from semi-leptonic beauty decays tend to be produced at a secondary vertex, displaced from the primary vertex, because of the long lifetime of $B$ hadrons. The signed impact parameter $\delta$ is calculated with respect to the beam position in the transverse plane (beam-spot). 
The beam-spot position was calculated every 2000 events as described elsewhere \cite{thesis:nicholass:2008}. The sign of $\delta$ is positive if the muon intercepts the axis of the associated jet within the jet hemisphere; otherwise $\delta$ is negative. 

The impact-parameter resolution in the MC was corrected \cite{thesis:miglioranzi:2006}, simultaneously taking into account the residual effects of multiple scattering and of the tracking resolution. The correction was extracted from inclusive-jet data by fitting the impact-parameter distribution with a double convolution of a Gaussian and a Breit-Wigner function. The widths of these functions were tuned taking into account the $p_T$ dependence of the size of the correction. 

Figure \ref{f:ptrelip}(b) shows the distribution of the reconstructed muon $\delta$ compared to predictions from the {\sc Pythia} MC model for beauty, charm and LF events which were corrected as described above. The $\delta$ distribution for LF events, which is symmetric and peaked at zero, has a finite width which reflects the impact-parameter resolution. Whereas for beauty events, and to a lesser extent for charm events, the $\delta$ distribution is asymmetric towards positive $\delta$. 

The fractions of beauty ($a_{b\bbar}$) and charm ($a_{c\cbar}$) events in the sample were obtained from a three-component fit, $f_{\mu}$, to the measured two-dimensional distributions of $p_{T}^{\rm \: rel}$ and $\delta$: 

\begin{equation}
f_{\mu}=a_{b\bbar}f_{\mu}^{b\bbar}+a_{c\cbar}f_{\mu}^{c\cbar}+(1-a_{b\bbar}-a_{c\cbar})f_{\mu}^{\rm LF},
\end{equation}

where $f_{\mu}^{b\bbar}$, $f_{\mu}^{c\cbar}$ and $f_{\mu}^{\rm LF}$ are the MC predicted shapes for beauty, charm and light flavour events. The fit used the minimum-${\chi}^{2}$ method and included MC statistical uncertainties. For differential cross sections the fit was repeated for each bin.

\begin{figure}[p]
\vfill
\begin{center}
\includegraphics[width=0.8\textwidth]{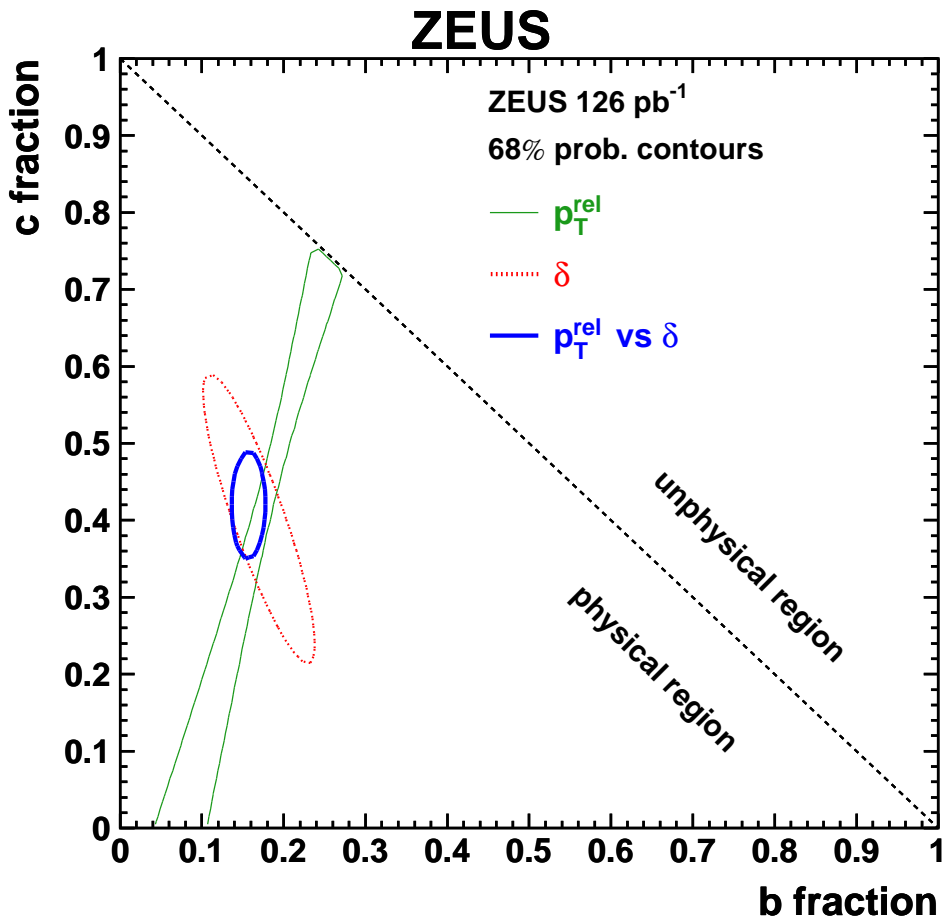}
\end{center}
\caption{
Contours of 68\% probability in the plane defined by the beauty and the charm fractions. The result of the $\chi^2$ fit to the two-dimensional $p_T^{\rm rel}$-$\delta$ distribution (thick solid line) and the one-dimensional distributions in $p_T^{\rm rel}$ (thin solid line) and $\delta$ (dotted line) are also shown. The diagonal line shows the boundary of the physical region in which the fractions of b, c and LF are positive.}
\label{f:ellipse}
\vfill
\end{figure}

As an illustration, Fig.~\ref{f:ellipse} shows the $68\%$ probability contours from the two-dimensional fit described above and from one-dimensional fits carried out using $p_{T}^{\rm \: rel}$ or $\delta$ alone for the data sample with $p_{T}^\mu>2.5 \gev$. Only statistical errors were taken into account. The two variables give complementary information. The $p_{T}^{\rm \: rel}$ fit alone is able to distinguish the $b$ component from charm and LF but not to separate these two background components, while the $\delta$ fit gives a good determination of the total heavy quark fraction. The $\delta$ fit also provides a strong anti-correlation between the fractions of beauty and charm\footnote{Due to the correlations between $p_{T}^{\rm \: rel}$ and $\delta$ the combined contour is not completely contained within the overlap of the two individual contours.}. In the previous analysis \cite{pr:d70:012008}, which used the $p_{T}^{\rm \: rel}$ method alone, the charm contribution was constrained to the charm cross section obtained from other measurements. This is not necessary here.

\section{Theoretical predictions and uncertainties}
\label{sec:theo}

The measured cross sections are compared to NLO QCD predictions based on the FMNR \cite{np:b412:225} program. The parton distribution functions used for the nominal prediction were GRVG-HO \cite{pr:d46:1973} for the photon and CTEQ5M \cite{epj:c12:375} for the proton. The $b$-quark mass was set to $m_{b}=4.75~\gev$, and the renormalisation and factorisation scales to the transverse mass, $\mu_{r}\!=\!\mu_{f}\!=m_{T}\!=\!\sqrt{\frac{1}{2}\left( (p_{T}^{b})^2+(p_{T}^{\bbar})^2 \right)+m_{b}^{2}}$, where $p_{T}^{b(\bbar)}$ is the transverse momentum of the $b$ ($\bbar$) quark in the laboratory frame. 
Jets were reconstructed by running the $k_{T}$ algorithm on the four-momenta of the $b$ and $\bbar$ quarks and of the third light parton (if present) generated by the program. The fragmentation of the $b$ quark into a $B$ hadron was simulated by rescaling the quark three-momentum (in the frame in which   $p^b_Z=-p^{\bbar}_Z$, obtained with a boost along $Z$) according to the Peterson \cite{pr:d27:105} fragmentation function with $\epsilon = 0.0035$. The muon momentum  was generated isotropically in the B-hadron rest frame from the decay spectrum given by {\sc Pythia} which is in good agreement with measurements made at B factories \cite{pl:b457:181,*pr:d67:031101}. 

The NLO cross sections, calculated for jets made of partons, were corrected for jet hadronisation effects to allow a direct comparison with the measured hadron-level cross sections. The correction factors, $C_{\rm had}$, were derived from the MC simulation as the ratio of the hadron-level to the parton-level MC cross section, where the parton level is defined as being the result of the parton-showering stage of the simulation. 

To evaluate the uncertainty on the NLO calculations, the $b$-quark mass and the renormalisation and factorisation scales were varied simultaneously to maximise the change, from $m_b=4.5~\rm{GeV}$ and  $\mu_{r}\!=\!\mu_{f}\!=m_{T}/2$ to $m_{b}=5.0~\rm{GeV}$ and $\mu_r\!=\!\mu_f=2m_{T}$, producing a variation in the cross section from $+34\%$ to $-22\%$. The effect on the cross section of a variation of the Peterson parameter $\epsilon$ and of a change of the fragmentation function from the Peterson to the Kartvelishvili parameterisation was found in a previous publication \cite{pr:d70:012008} to be of the order of $3\%$. The effects of using different sets of parton densities and of a variation of the strong coupling constant were found to be within $\pm 4\%$. These effects are negligible with respect to that of a variation of the $b$-quark mass and the renormalisation and factorisation scales and are therefore not included. The uncertainty due to the hadronisation correction was also found to be negligible with respect to the dominant uncertainty.

The measured cross sections are also compared to the predictions of the {\sc Pythia\,6.2} MC model scaled to the data. The predictions of {\sc Pythia} were obtained \cite{epj:c17:137} by mixing flavour-creation processes ($\gamma g \rightarrow b \bbar$, $g g \rightarrow b \bbar$, $q \qbar \rightarrow b \bbar$) calculated using massive matrix elements and the flavour-excitation (FE) processes ($b g  \rightarrow b g$,  $b q  \rightarrow b q$), in which a heavy quark is extracted from the photon or proton parton density. The FE processes contribute about $27\%$ to the total $b\bbar$ cross section. The small ($\sim 5\%$)\cite{pr:d70:012008} contribution from final-state gluon splitting in parton showers ($g  \rightarrow b \bbar$) was not included. The parton density CTEQ5L \cite{epj:c12:375} was used for the proton and GRVG-LO \cite{pr:d46:1973} for the photon; the $b$-quark mass was set to $4.75$ GeV and the $b$-quark string fragmentation was performed according to the Peterson function with $\epsilon=0.0041$ \cite{pr:d27:105}.

\section{Systematic uncertainties}
\label{sec:sys}

The main experimental uncertainties are calculated as follows (the resulting uncertainty on the total cross section is given in parentheses)\cite{thesis:boutle:2009}:

\begin{itemize}
\item the muon acceptance, including the efficiency of the muon chambers, of the reconstruction and of the B/RMUON matching to central tracks, is known to about $7\%$. An independent dimuon sample was analysed to determine this uncertainty based on a method \cite{thesis:turcato:2002} which has been repeated here ($\pm 7\%$);
\item the error due to the uncertainty of the energy scale of the CAL was evaluated by varying the energy of the jets and the inelasticity $y_{\rm JB}$ in the MC by $\pm 3\%$ ($\pm 4\%$);
\item the efficiency of finding a track with 4 MVD hits was measured in the data and in the MC. The ratio of the measured efficiencies was applied as a correction to the acceptance. The uncertainty on this ratio was included in the systematic uncertainty ($\pm 3\%$);
\item the efficiency of the dijet trigger in the MC was corrected so that it reproduced the efficiency as measured in the data. The systematic uncertainty due to this correction was negligible;
\item the MC $\eta^{\mu}$ distribution was reweighted in order to account for the differences (see Fig.~\ref{f:f1}) between data and MC ($<1\%$);
\item the uncertainty on the size of the correction to the shape of the impact-parameter distribution for the MC samples described in Section~\ref{sec:cros} was evaluated by varying the widths of the Gaussian and Breit-Wigner distributions used in the correction function by $+ 20\%$ and $- 10\%$ of their nominal values. These variations are such that the global MC distribution still provides a good description of the data ($^{+6\%}_{-10\%}$);
\item the uncertainty on the $p_T^{\rm rel}$ shape of the LF and charm background was evaluated by:
  \begin{itemize}
  \item varying the correction applied to the LF background by $\pm 20\%$ of its nominal value ($\pm 2\%$);
  \item varying the  $p_T^{\rm rel}$ shape of the charm component by removing or doubling the correction ($\pm 4\%$);
  \end{itemize}
\item the contribution of flavour-excitation events in {\sc Pythia} was varied\footnote{The comparison between MC prediction and data for $d\sigma / dx_{\gamma}^{jj}$, see Section~\ref{sec:res}, is still satisfactory for these variations.} by $+\,100\%/-~50\%$ and simultaneously the contribution of $g g \rightarrow b \bbar$, $q \qbar \rightarrow b \bbar$ events was varied by $-\,50\%/+\,100\%$ ($\pm 4\%$); the contribution of $\gamma g \rightarrow b \bbar$ processes in {\sc Pythia} was decreased by $20\%$ and all other processes were increased by $+100\%$ ($\pm 2\%$).

\end{itemize}

The total systematic uncertainty was obtained by adding the above contributions in quadrature. A 2.6\% overall normalisation uncertainty associated with the luminosity measurement was not included in the systematic uncertainty on the differential cross sections.

\section{Results}
\label{sec:res}

\begin{figure}[p]
\vfill
\begin{center}
\includegraphics[width=0.9\textwidth]{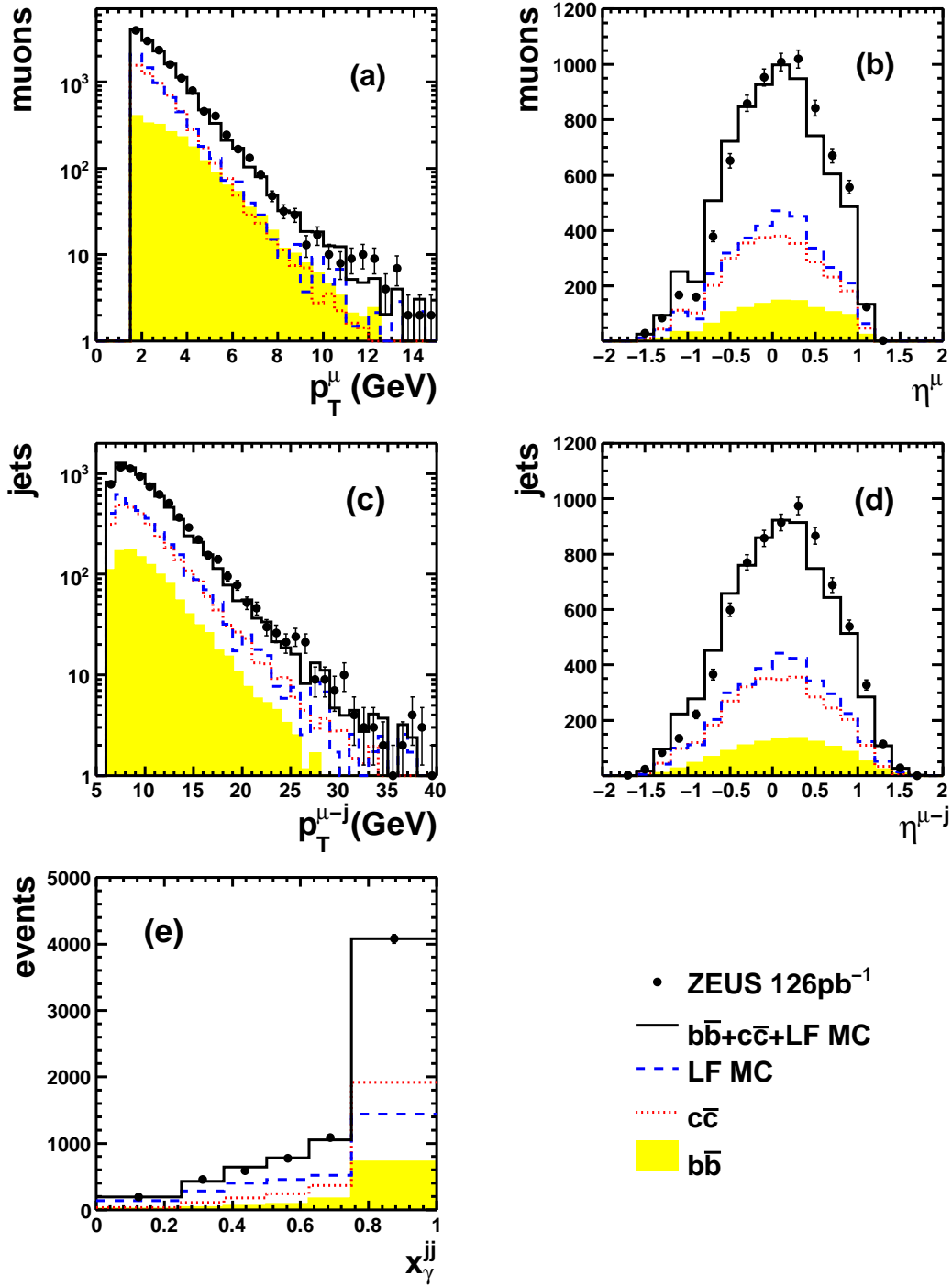}
\end{center}
\caption{
Distributions of (a) $p_{T}^{\mu}$, (b) $\eta^{\mu}$, (c) $p_{T}^{\mu\mbox{-}j}$, (d) $\eta^{\mu\mbox{-}j}$ and (e) $x_{\gamma}^{jj}$. The data are compared to a mixture of beauty (shaded histogram), charm (dotted line) and light flavour(dashed line) {\sc Pythia} MC predictions, combined according to the fractions given by the two-dimensional $p_{T}^{\rm \:rel}$-$\delta$ fit. The total MC distribution is shown as the solid line. The kinematic region is restricted to $p_{T}^{\mu} > 1.5~\rm GeV$ ($p_{T}^{\mu} > 2.5~\rm GeV$) for (a)((b)-(e)).}

\label{f:f1}
\vfill
\end{figure}

Figure \ref{f:f1} shows the distributions of the kinematic variables of the muon $p_{T}^{\mu}$ and $\eta^{\mu}$ as well as those for the jet associated with the muon $p_{T}^{\mu-j}$ and $\eta^{\mu-j}$. The fraction $x_{\gamma}^{jj}$ of the total hadronic $E-p_{Z}$ carried by the two highest-$p_{T}$ jets\footnote{$x_{\gamma}^{jj}$ is the massive-jets analogue of the $x_{\gamma}^{\rm obs}$ variable used for massless jets in other ZEUS publications\cite{np:b729:492}.} is given by:

\begin{equation}
x_{\gamma}^{jj}=\frac{\sum_{j=1,2} (E^{j}-p^{j}_Z)}{E-p_Z}.
\label{eq:xgamma}
\end{equation}

The distribution of $x_{\gamma}^{jj}$ is also shown in Fig.~\ref{f:f1}.
The data are compared in shape to the MC simulations in which the relative contributions of beauty, charm and LF were mixed according to the fractions measured in this analysis as described in Section \ref{sec:cros}. The comparison shows that the main features of the dijet-plus-muon sample are reasonably well reproduced by this MC mixture.

Total and differential visible cross sections have been measured for final states with at least one muon and two jets in the following kinematic region:

\begin{itemize}
\item $Q^2 < 1 \rm \,GeV^{2}$ and $0.2 < y < 0.8;$
\item $p_{T}^{j1,j2} > 7,6 \, \rm GeV$ and $|\eta^{j1,j2}| < 2.5;$ the jets are defined as hadron-level jets using the $k_T$ algorithm. For the purposes of jet-finding, $B$ hadrons are treated as stable particles;
\item $p_{T}^\mu > 2.5 \, \rm GeV$ ($p_{T}^\mu > 1.5 \, \rm GeV$ for $d\sigma / dp_{T}^\mu$) and $-1.6 < \eta^{\mu} < 1.3;$
\item at least one muon is associated with a jet with $p_{T}^{j} > 6 \, \rm GeV$. The muon is associated with the jet if it is the decay product of a $B$ hadron contained in the jet, according to the $k_T$ algorithm. Muons coming from both direct ($b \rightarrow \mu$) and indirect ($b \rightarrow c,\cbar,J/\Psi,\Psi' \rightarrow \mu$) decays are considered to be part of the signal.
\end{itemize}

The total visible cross section is

\begin{equation}
\sigma(ep \rightarrow e b\bbar X \rightarrow e jj \mu X') = 38.6 \pm 3.5 (\rm stat.) _{-4.9}^{+4.6} (\rm syst.) ~ \rm pb.
\end{equation}

This result is compared to the NLO QCD calculation described in Section \ref{sec:theo}. The prediction for the total visible cross section is

\begin{equation}
\sigma(ep \rightarrow e b\bbar X \rightarrow e jj \mu X') = 37.0 _{-7.5}^{+11.9} ~ \rm pb,
\end{equation}

in excellent agreement with the data.

\begin{figure}[p]
\vfill
\begin{center}
\includegraphics[width=\textwidth]{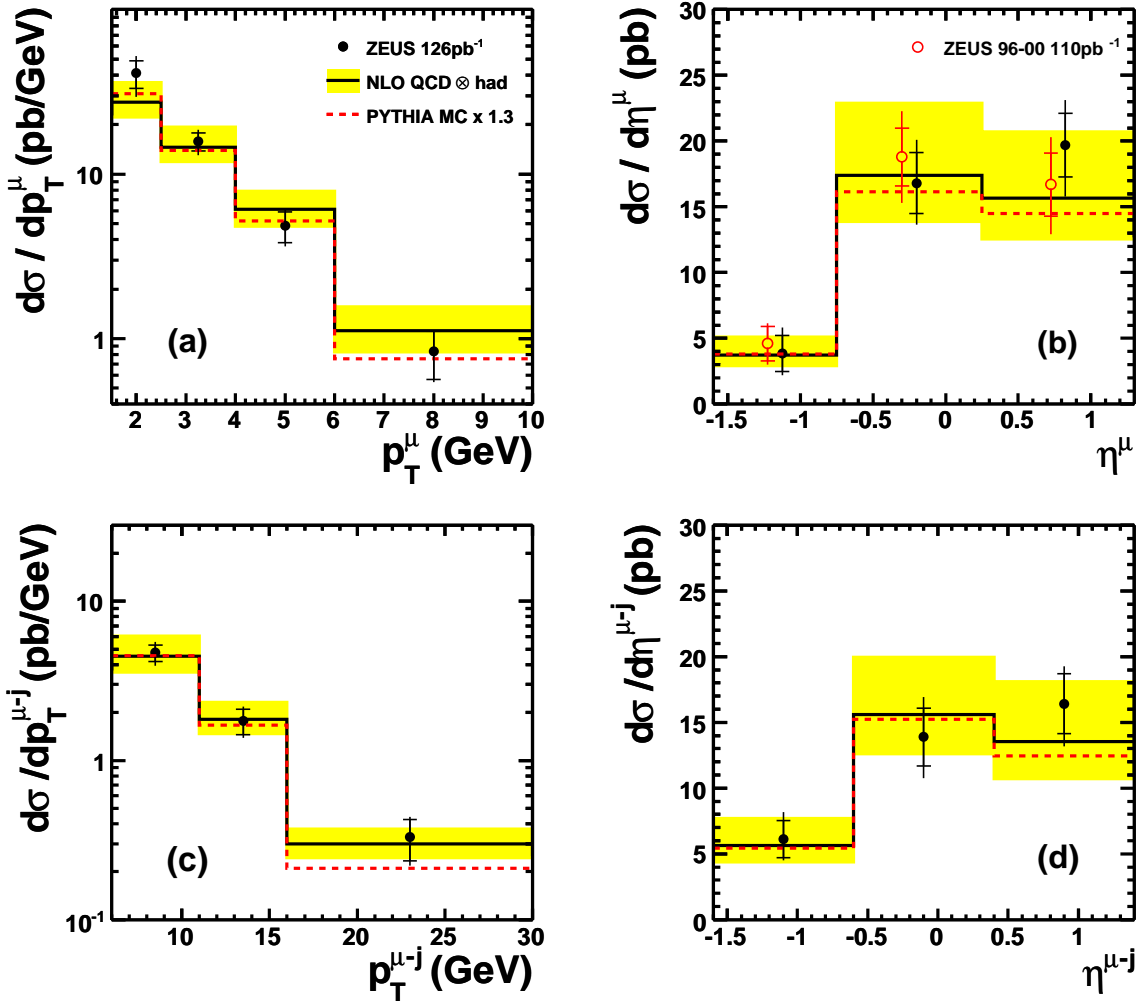}
\end{center}
\caption{
Differential cross section as a function (a) $p_{T}^{\mu}$, (b) $\eta^{\mu}$, (c) $p_{T}^{\mu\mbox{-}j}$ and (d) $\eta^{\mu\mbox{-}j}$ for $Q^2<1\gev^2$, $0.2<y<0.8$, $p_T^{j_1,j_2}>7,6 \gev$, $|\eta_{j_1,j_2}| <2.5$, and $-1.6<\eta^{\mu}<1.3$. For the $p_{T}^{\mu}$ cross section, the kinematic region is defined as $p_{T}^{\mu} > 1.5~\rm GeV$ and as $p_{T}^{\mu} > 2.5~\rm GeV$ for all other cross sections. The filled circles show the results from this analysis and the open circles show the results from the previous ZEUS measurement. The inner error bars are statistical uncertainties while the external bars show the statistical and systematic uncertainties added in quadrature. The band represents the NLO QCD predictions with their uncertainties. The {\sc Pythia} MC predictions are also shown (dashed line).}
\label{fig:muon}
\vfill
\end{figure}

Figure \ref{fig:muon} and Table \ref{tab-muon} show the visible differential cross sections as a function of the muon transverse momentum and pseudorapidity. Also shown in Fig. \ref{fig:muon} and in Table \ref{tab-mujets} are the visible differential cross sections measured as a function of the transverse momentum of the jet associated with the muon  $p_{T}^{\mu-j}$, and as a function of its pseudorapidity, $\eta^{\mu-j}$. The visible cross section as a function of $p_{T}^\mu$ is measured in the range $p_{T}^\mu > 1.5~ \rm GeV$, while the other cross sections are measured for $p_{T}^\mu > 2.5~ \rm GeV$. 
The NLO QCD predictions describe the data well and the {\sc Pythia} MC also gives a good description of the shapes.

The visible differential cross section as a function of $\eta^\mu$ is also compared with a previous ZEUS measurement \cite{pr:d70:012008}, which used the $p_{T}^{\rm \: rel}$ method to extract the beauty fraction. The two measurements agree well\footnote{The measurement presented by H1 \cite{epj:c41:453} refers to a slightly different definition of the cross section and therefore cannot be compared to directly. However a qualitative comparison does not confirm their observation of an excess at low $p_{T}^\mu$.}.

\begin{figure}[p]
\vfill
\begin{center}
\includegraphics[width=\textwidth]{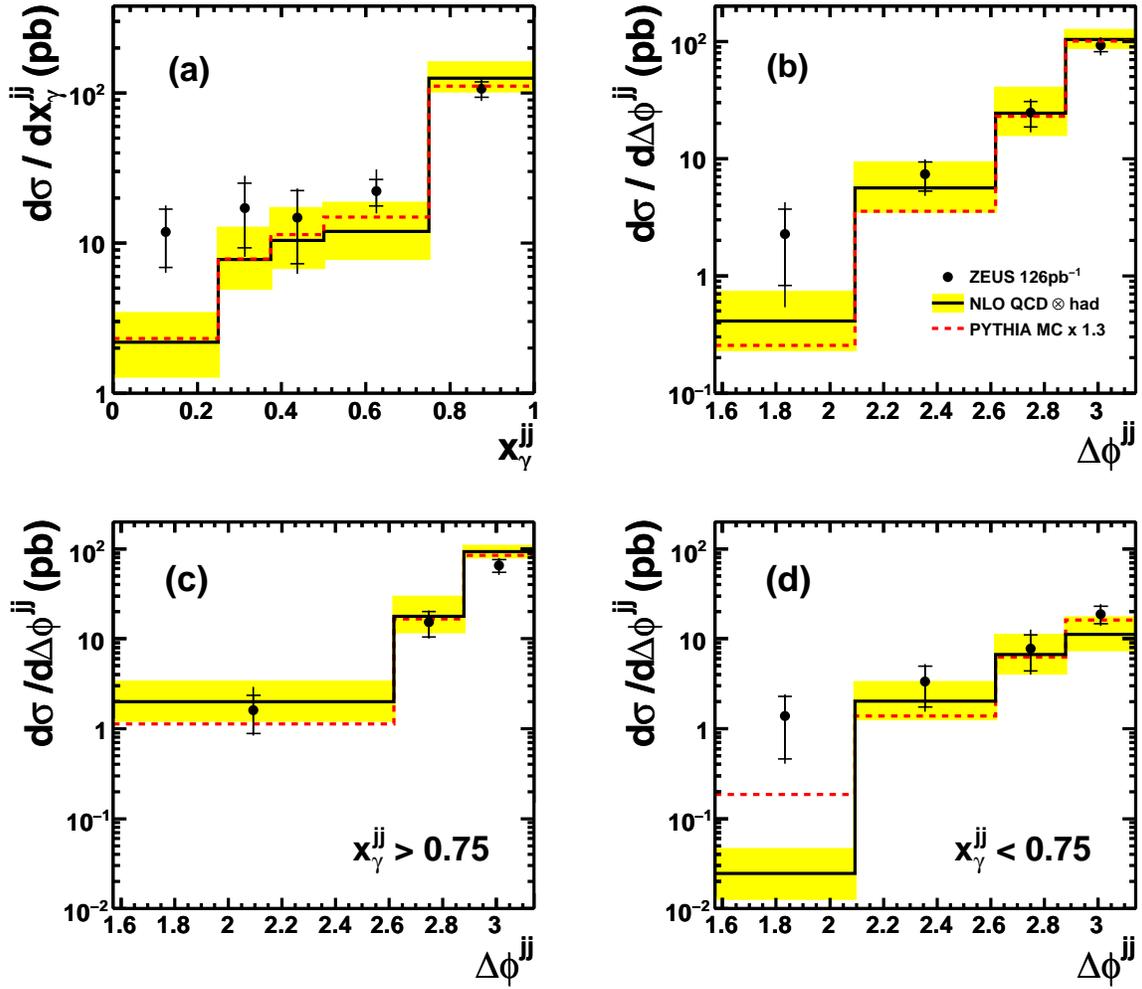}
\end{center}
\caption{
Differential cross sections as a function of (a) $x_\gamma^{jj}$ and (b) $\Delta\phi^{jj}$ of the jet-jet system and $\Delta\phi^{jj}$ for (c) direct- and (d) resolved-enriched samples for $Q^2<1\gev^2$, $0.2<y<0.8$, $p_T^{j_1,j_2}>7,6 \gev$, $\eta_{j_1,j_2}<2.5$, $p_{T}^{\mu}>2.5~\rm GeV$ and $-1.6<\eta^{\mu}<1.3$. The inner error bars are statistical uncertainties while the external bars show the statistical and systematic uncertainties added in quadrature. The band represents the NLO QCD predictions with their uncertainties. The {\sc Pythia} MC predictions are also shown (dashed line).}
\label{fig:dijet}
\vfill
\end{figure}

Figure \ref{fig:dijet}(a) and Table \ref{tab-xgam} show the visible dijet cross section as a function of $x_{\gamma}^{jj}$ (Eq.~\ref{eq:xgamma}). The $x_{\gamma}^{jj}$ variable corresponds at LO to the fraction of the exchanged-photon momentum entering the hard scattering process. In photoproduction, events can be classified into two types of process in LO QCD. In direct processes, the photon couples as a point-like object in the hard scatter. In resolved processes, the photon acts as a source of incoming partons, one of which takes part in the hard scatter. The $x_{\gamma}^{jj}$ variable provides a tool to measure the relative importance of direct processes, which gives a peak at $x_{\gamma}^{jj} \sim 1$, and of resolved processes, which are distributed over the whole $x_{\gamma}^{jj}$ range. The dominant contribution to the visible cross section comes from the high-$x_{\gamma}^{jj}$ peak but a low-$x_{\gamma}^{jj}$ component is also apparent. The NLO QCD prediction describes the measured visible cross section well. {\sc Pythia} also gives a good description of the shape of the distribution.

Dijet angular correlations are particularly sensitive to higher-order effects and are therefore suitable to test the limitations of fixed-order perturbative QCD calculations.  The dijet variable measured, $\Delta\phi^{jj}$, was reconstructed from the two highest-$p_{T}$ jets as:

\begin{equation}
\Delta\phi^{jj} = |\phi^{j1}-\phi^{j2}|~.
\label{eqdphijj}
\end{equation}

In the FMNR program, at LO the differential cross section as a function of $\Delta\phi^{jj}$ is a delta function peaked at $\pi$. At NLO, exclusive three-jet production populates the region $\frac{2}{3}\pi < \Delta\phi^{jj} < \pi $, whilst smaller values of $\Delta\phi^{jj}$ require additional radiation such as a fourth jet in the event. An NLO QCD calculation can produce values of $\Delta\phi^{jj} < \frac{2}{3}\pi$ when the highest-$p_{T}$ jet is not in the accepted kinematic region.

The visible differential cross section as a function of $\Delta\phi^{jj}$ is shown in Fig.~\ref{fig:dijet}(b) and Table \ref{tab-dijet}. The NLO QCD predictions describe the data well. Visible cross sections as a function of $\Delta\phi^{jj}$ have also been measured separately for direct-enriched ($x_{\gamma}^{jj} > 0.75$) and resolved-enriched ($x_{\gamma}^{jj} < 0.75$) samples (Fig.~\ref{fig:dijet}(c) and (d)). The cross sections are well described by the NLO QCD prediction for $x_{\gamma}^{jj} > 0.75$ and for $x_{\gamma}^{jj} < 0.75$. The {\sc Pythia} MC gives an equally good description of the shape of the distributions.

\section{Conclusions}
\label{sec:con}

Beauty production identified through semi-leptonic decay into muons has been measured with the ZEUS detector in the kinematic range defined as: $Q^2 < 1 \, \rm GeV^{2}$; $0.2 < y < 0.8$; $p_{T}^{j1,j2} > 7,6 \, \rm GeV$; $|\eta^{j1,j2}| < 2.5$; $p_{T}^\mu > 2.5 \, \rm GeV$; $-1.6 < \eta^{\mu} < 1.3$ with at least one muon being associated with a jet with $p_{T}^{j} > 6 \, \rm GeV$. Lifetime information was combined with the muon $p_T^{\rm rel}$ method to extract the fraction of beauty events in the data sample. Unlike the previous analysis, which used the $p_{T}^{\rm \: rel}$ method alone, it was not necessary to constrain the charm contribution to the charm cross section obtained from other measurements. The extracted charm contribution is consistent with expectation.

The total visible cross section was measured as well as visible differential cross sections as a function of the transverse momenta and pseudorapidities of the muon and of the jet associated with the muon. The $\eta^{\mu}$ cross section was compared to the previous measurement. This analysis confirms the previous result with similar statistical precision and different sources of systematic uncertainty. Also, it was possible to measure the cross section as a function of the muon transverse momentum to $p_{T}^\mu > 1.5 \, \rm GeV$, a lower $p_{T}^\mu$ than in the previous muon-jet analysis. The $p_{T}^\mu$ cross section agrees well with the NLO QCD prediction and does not confirm the excess observed by H1 at low $p_{T}^\mu$.

All results were compared to the {\sc Pythia} MC model and to an NLO QCD prediction. The NLO QCD prediction describes the data well. The {\sc Pythia} MC model also provides a good description of the shape of the distributions. 

Beauty dijet angular-correlation cross sections were also measured. Separate measurements in direct-enriched and resolved-enriched regions were presented. The NLO QCD prediction describes the measured cross sections well. 

\section{Acknowledgements}
\label{sec:ack}

We appreciate the contributions to the construction and maintenance of the ZEUS detector of many people who are not listed as authors. The HERA machine group and the DESY computing staff are especially acknowledged for their success in providing excellent operation of the collider and the data-analysis environment. We thank the DESY directorate for their strong support and encouragement. 

{
\def\bibname{\Large\bf References}
\def\refname{\Large\bf References}
\pagestyle{plain}
{\raggedright
\bibliography{DESY-08-210.bbl}}
}
\vfill\eject


\begin{table}
\begin{center}
\begin{tabular}{|c|c|c|}
\hline
$p_{T}^{\mu}$ range & $d\sigma/dp_{T}^{\mu}\pm{\rm stat.}\pm{\rm syst.}$   & $C_{\rm had}$ \\
 (GeV)  & (pb/GeV) & \\ 
\hline
\begin{tabular}{r@{}l@{}r@{}}
1.5 &,& 2.5  \\
2.5 &,& 4.0  \\
4.0 &,& 6.0  \\
6.0 &,& ~10.0 \\
\end{tabular}

& 

\begin{tabular}{r@{}l@{}}
$41.05 \pm 7.74$& $ ^{+8.26}_{-8.51}$ \\
$15.78 \pm 1.96$& $ ^{+2.03}_{-1.98}$ \\
$4.87 \pm 1.03$& $ ^{+0.69}_{-0.67}$ \\
$0.84 \pm 0.27$& $ ^{+0.11}_{-0.11}$ \\
\end{tabular}

& 

\begin{tabular}{r@{}}
$0.87$ \\
$0.93$ \\
$0.98$ \\
$1.01$ \\
\end{tabular}

\\

\hline \hline
$\eta^{\mu}$ range  & $d\sigma/d\eta^{\mu}\pm{\rm stat.}\pm{\rm syst.}$  & $C_{\rm had}$ \\
  & (pb) & \\ 
\hline

\begin{tabular}{r@{}@{}l@{}r@{}}
$-$1.60&,& $-$0.75 \\
$-$0.75&,&0.25 \\
0.25&,&1.30 \\
\end{tabular}

& 

\begin{tabular}{r@{}}
$3.86 \pm 1.37 ^{+1.40}_{-0.92}$ \\
$16.81 \pm 2.30 ^{+2.34}_{-2.15}$ \\
$19.70 \pm 2.43 ^{+2.43}_{-3.09}$ \\
\end{tabular}

& 

\begin{tabular}{r@{}}
$0.83$ \\
$0.89$ \\
$0.92$ \\
\end{tabular}

\\
\hline
\end{tabular}
\caption{
Differential muon cross section as a function of
$p_{T}^{\mu}$ and $\eta^{\mu}$.
For further details see text.
The multiplicative hadronisation correction, $C_{\rm had}$,
applied to the NLO prediction is shown in the last column.
}
  \label{tab-muon}
\end{center}
\end{table}

\begin{table}
\begin{center}
\begin{tabular}{|c|c|c|} \hline
$p_{T}^{\mu\mbox{-}j}$ range & $d\sigma/dp_{T}^{\mu\mbox{-}j}\pm{\rm stat.}\pm{\rm syst.}$  & $C_{\rm had}$ \\
(GeV) & (pb/GeV) & \\ \hline

\begin{tabular}{r@{}l@{}r@{}}
6&,& ~11 \\
11&,& ~16 \\
16&,& ~30 \\
\end{tabular}

&

\begin{tabular}{r@{}}
$4.74 \pm 0.57 ^{+0.60}_{-0.59}$ \\
$1.78 \pm 0.32 ^{+0.24}_{-0.22}$ \\
$0.33 \pm 0.10 ^{+0.05}_{-0.05}$ \\
\end{tabular}

&

\begin{tabular}{r@{}}
$0.89$ \\
$0.89$ \\
$0.92$ \\
\end{tabular}

\\

\hline \hline
$\eta^{\mu\mbox{-}j}$ range & $d\sigma/d\eta^{\mu\mbox{-}j}\pm{\rm stat.}\pm{\rm syst.}$  & $C_{\rm had}$ \\ 
 & (pb) &  \\ \hline
\begin{tabular}{r@{}l@{}r@{}}
$-$1.6&,& $-$0.6 \\
$-$0.6&,& 0.4 \\
 0.4&,& 1.4 \\
\end{tabular}

&

\begin{tabular}{r@{}l@{}}
$6.13 \pm 1.41$& $ ^{+1.50}_{-0.82}$ \\
$13.89 \pm 2.20$& $ ^{+2.08}_{-2.21}$ \\
$16.42 \pm 2.29$& $ ^{+1.70}_{-2.29}$ \\
\end{tabular}

&

\begin{tabular}{r@{}}
$0.77$ \\
$0.84$ \\
$0.99$ \\
\end{tabular}

\\

\hline
\end{tabular}
\caption{\label{tab-mujets}
Differential cross section for jets associated with
a muon as a function of $p_{T}^{\mu{\mbox{-}j}}$
and $\eta^{\mu{\mbox{-}j}}$.
For further details see text.
}
\end{center}
\end{table}

\begin{table}
\begin{center}
\begin{tabular}{|c|c|c|}\hline
$x_{\gamma}^{jj}$ range  & $d\sigma/dx_{\gamma}^{jj}\pm{\rm stat.}\pm{\rm syst.}$& $C_{\rm had}$ \\
& (pb) & \\ \hline

\begin{tabular}{r@{}}
0.000, 0.250 \\
0.250, 0.375 \\
0.375, 0.500 \\
0.500, 0.750 \\
0.750, 1.000 \\
\end{tabular}

&

\begin{tabular}{r@{}r@{}r@{}l@{}}
$11.85 $~&$ \pm $&$ 4.96 $&~$ ^{+3.32}_{-2.40}$ \\
$17.17 $~&$ \pm $&$ 7.89 $&~$ ^{+7.69}_{-4.47}$ \\
$14.81 $~&$ \pm $&$ 7.56 $&~$ ^{+3.30}_{-4.06}$ \\
$22.19 $~&$ \pm $&$ 4.48 $&~$ ^{+7.47}_{-4.51}$ \\
$106.63 $~&$ \pm $&$ 12.63 $&~$ ^{+11.82}_{-12.74}$ \\
\end{tabular}

&

\begin{tabular}{r@{}}
$0.69$ \\
$0.78$ \\
$0.86$ \\
$0.86$ \\
$0.92$ \\
\end{tabular}

\\

\hline 
\end{tabular}
\caption{\label{tab-xgam}
Differential cross section as a function of
$x_{\gamma}^{jj}$.
For further details see text.}
\end{center}
\end{table}

\begin{table}
\begin{center}
\begin{tabular}{|c|c|c|}\hline
$\Delta\phi^{jj}$ range  & $d\sigma/d\Delta\phi^{jj}\pm{\rm stat.}\pm{\rm syst.}$& $C_{had}$ \\
& (pb) & \\ \hline

\begin{tabular}{r@{}l@{}r@{}}
${\frac{6\pi}{12}}$ &,& ${\frac{8\pi}{12}}$ \\
${\frac{8\pi}{12}}$  &,&  ${\frac{10\pi}{12}}$ \\
${\frac{10\pi}{12}}$  &,&  ${\frac{11\pi}{12}}$ \\
${\frac{11\pi}{12}}$  &,&  ${\frac{12\pi}{12}}$ \\
\end{tabular}

&

\begin{tabular}{r@{}r@{}r@{}l@{}}
$2.26 $~&$ \pm $&$ 1.44 $&~$ ^{+1.34}_{-0.96}$ \\
$7.35 $~&$ \pm $&$ 2.06 $&~$ ^{+1.47}_{-1.45}$ \\
$24.70 $~&$ \pm $&$ 6.04 $&~$ ^{+4.66}_{-5.12}$ \\
$92.91 $~&$ \pm $&$ 11.10 $&~$ ^{+10.46}_{-12.82}$ \\
\end{tabular}

&

\begin{tabular}{r@{}}
$0.80$ \\
$0.79$ \\
$0.86$ \\
$0.92$ \\
\end{tabular}

\\

\hline
\end{tabular}

$x_{\gamma}^{ jj} > 0.75$\\
\begin{tabular}{|c|c|c|}\hline 
$\Delta\phi^{ jj}$ range  & $d\sigma/d\Delta\phi^{ jj}\pm{\rm stat.}\pm{\rm syst.}$& $C_{had}$ \\
& (pb) & \\ \hline
\begin{tabular}{r@{}l@{}r@{}}
${\frac{6\pi}{12}}$ &,& ${\frac{10\pi}{12}}$ \\
${\frac{10\pi}{12}}$ &,& $ {\frac{11\pi}{12}}$ \\
${\frac{11\pi}{12}}$ &,& $ {\frac{12\pi}{12}}$ \\
\end{tabular}

&

\begin{tabular}{r@{}r@{}r@{}l@{}}
$1.62 $~&$ \pm $&$ 0.73 $&~$ ^{+1.08}_{-0.27}$ \\
$15.27 $~&$ \pm $&$ 4.75 $&~$ ^{+2.50}_{-2.21}$ \\
$65.69 $~&$ \pm $&$ 10.66 $&~$ ^{+8.14}_{-9.18}$ \\
\end{tabular}

&

\begin{tabular}{r@{}}
$0.82$ \\
$0.87$ \\
$0.93$ \\
\end{tabular}

\\

\hline 
\end{tabular}

 $x_{\gamma}^{ jj} < 0.75$\\
\begin{tabular}{|c|c|c|}\hline 
$\Delta\phi^{ jj}$ range  & $d\sigma/d\Delta\phi^{ jj}\pm{\rm stat.}\pm{\rm syst.}$& $C_{\rm had}$ \\
& (pb) & \\ \hline
\begin{tabular}{r@{}l@{}r@{}}
${\frac{6\pi}{12}}$ &,& $ {\frac{8\pi}{12}}$ \\
${\frac{8\pi}{12}}$ &,& $ {\frac{10\pi}{12}}$ \\
${\frac{10\pi}{12}}$ &,& $ {\frac{11\pi}{12}}$ \\
${\frac{11\pi}{12}}$ &,& $ {\frac{12\pi}{12}}$ \\
\end{tabular}

&

\begin{tabular}{r@{}r@{}r@{}l@{}}
$1.38 $~&$ \pm $&$ 0.92 $&~$ ^{+0.46}_{-0.31}$ \\
$3.36 $~&$ \pm $&$ 1.60 $&~$ ^{+0.97}_{-0.87}$ \\
$7.75 $~&$ \pm $&$ 3.37 $&~$ ^{+3.67}_{-1.62}$ \\
$18.84 $~&$ \pm $&$ 4.17 $&~$ ^{+3.71}_{-2.59}$ \\
\end{tabular}

&

\begin{tabular}{r@{}}
$0.75$ \\
$0.76$ \\
$0.84$ \\
$0.84$ \\
\end{tabular}

\\

\hline 
\end{tabular}

\caption{\label{tab-dijet}
Differential muon cross section as a function of
$\Delta\phi^{jj}$ for all $x_{\gamma}^{jj}$ and for $x_{\gamma}^{ jj} >(<) 0.75$.
For further details see text.}
\end{center}
\end{table}


%
%
\end{document}